\documentclass[10pt,journal,compsoc]{IEEEtran}

\usepackage[utf8]{inputenc}
\usepackage{amsmath}
\usepackage{amsfonts}
\usepackage{amssymb}
\usepackage{multirow}
\usepackage{float}

\usepackage{graphics}
\usepackage{balance}
\usepackage{dsfont}
\usepackage{caption}
\usepackage{subcaption}
\usepackage{algorithm}
\usepackage{algorithmic}

\usepackage{latexsym}
\usepackage{soul}
\usepackage{url}
\usepackage{graphicx}
\usepackage{booktabs}
\usepackage{flushend}
\usepackage[T1]{fontenc}
\usepackage{algorithm}
\usepackage{algorithmic}
\usepackage{ulem}
\newcommand{\ie}{\textit{i.e.}}
\newcommand{\eg}{\textit{e.g.}}

\newcommand{\ours}{{QASS}}

\newcommand{\todo}[1]{\{\textcolor{blue}{\textbf{TODO}}\}}

\ifCLASSOPTIONcompsoc
  \usepackage[nocompress]{cite}
\else
  \usepackage{cite}
\fi

\ifCLASSINFOpdf
\else
\fi
\begin{document}
\title{Query-oriented Data Augmentation for Session Search}

\author{Haonan~Chen,
        Zhicheng~Dou,~\IEEEmembership{Member,~IEEE,}
        Yutao~Zhu,
        Ji-Rong~Wen,~\IEEEmembership{Senior~Member,~IEEE}
\IEEEcompsocitemizethanks{\IEEEcompsocthanksitem H. Chen, Z. Dou, Y. Zhu and J.-R Wen are with Gaoling School of Artificial Intelligence, Renmin University of China, Engineering Research Center of Next-Generation Intelligent Search and Recommendation, Ministry of Education, and Beijing Key Laboratory of Big Data Management and Analysis Methods, Beijing 100872, P.R. China.\protect\\
E-mail: hnchen, dou, ytzhu, jrwen@ruc.edu.cn.\protect\\
(Corresponding author: Zhicheng Dou)}}

\markboth{Journal of \LaTeX\ Class Files,~Vol.~14, No.~8, August~2015}%
{Shell \MakeLowercase{\textit{et al.}}: Bare Demo of IEEEtran.cls for Computer Society Journals}

\IEEEtitleabstractindextext{%
\begin{abstract}
Modeling contextual information in a search session has drawn more and more attention when understanding complex user intents. Recent methods are all data-driven, \ie, they train different models on large-scale search log data to identify the relevance between search contexts and candidate documents. 
The common training paradigm is to pair the search context with different candidate documents and train the model to rank the clicked documents higher than the unclicked ones. However, this paradigm neglects the symmetric nature of the relevance between the session context and document, \ie, the clicked documents can also be paired with different search contexts when training. 
In this work, we propose query-oriented data augmentation to enrich search logs and empower the modeling. We generate supplemental training pairs by altering the most important part of a search context, \ie, the current query, and train our model to rank the generated sequence along with the original sequence. This approach enables models to learn that the relevance of a document may vary as the session context changes, leading to a better understanding of users' search patterns. We develop several strategies to alter the current query, resulting in new training data with varying degrees of difficulty. Through experimentation on two extensive public search logs, we have successfully demonstrated the effectiveness of our model.

\end{abstract}
}

\maketitle

\IEEEdisplaynontitleabstractindextext
\IEEEpeerreviewmaketitle

\section{Introduction} \label{sec:intro}

\IEEEPARstart{A}s search intents continue to grow in complexity, and the search behavior of users has undergone significant changes, transitioning from the use of single queries to engaging in multiple interactions with search engines.
These interactions, including the queries issued and the documents clicked, form a search session. It has been shown that the information of a search session's context can facilitate the comprehension of the actual search intent.

\begin{figure}[tbp]
\includegraphics[width=\linewidth]{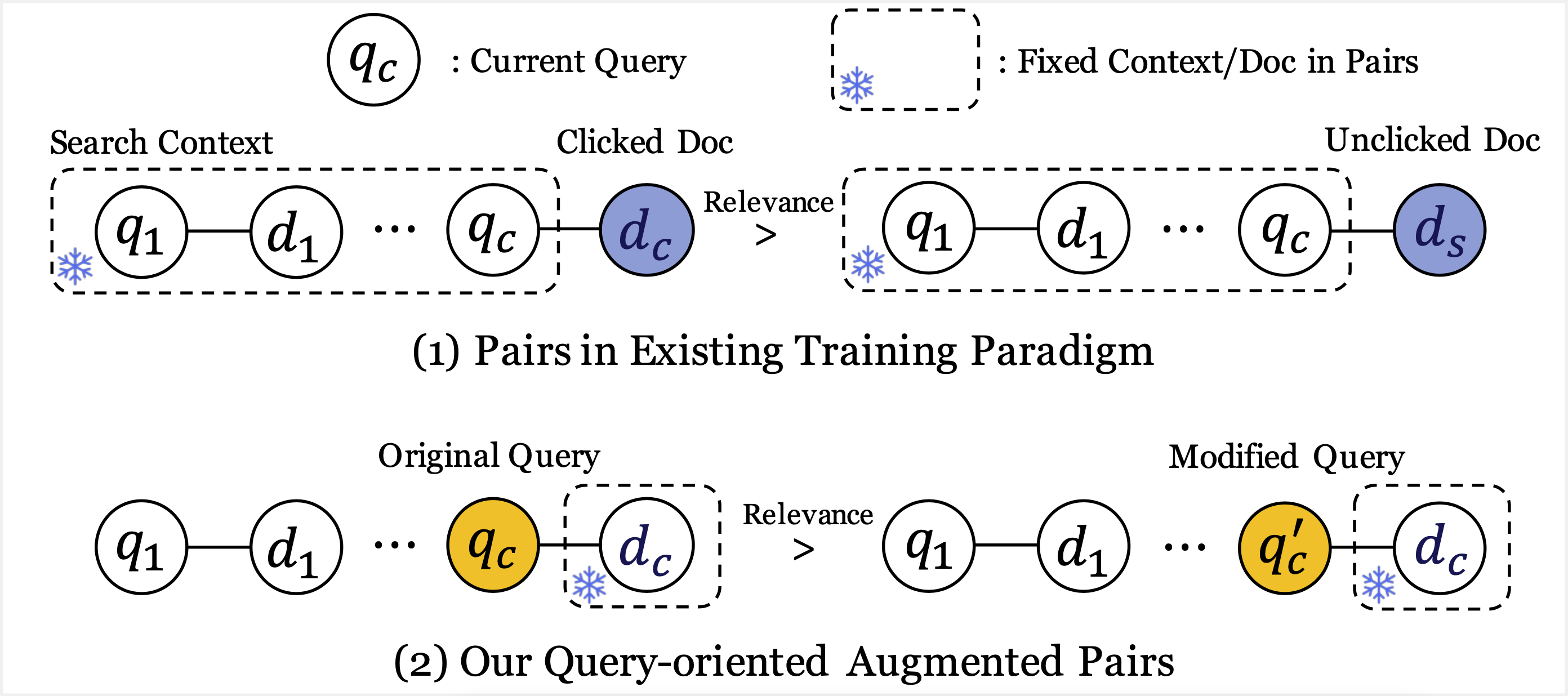}
\caption{An illustration of our augmented training pairs. The existing training paradigm constructs training samples by pairing different candidate documents with a fixed search context, while we pair fixed clicked document with the original search context and the one with the modified current query.}
\label{fig:cda_example} 
\end{figure}

These years, many neural approaches have been proposed to model the session sequence and rank the candidate documents. 
These models aim to extract valuable information from the search context to predict users' search intents.
For instance, some models employed recurrent neural networks (RNNs) to model search behaviors sequentially within a session~\cite{mmtensor,cars}.
Most recently, sophisticated pre-trained language models (PLMs), \eg, BERT~\cite{bert} and BART~\cite{ase}, have also been applied to model contextual user behaviors and calculate ranking scores.
{All these models are trained on the search log data with each sample} organized as a $\langle$search context, candidate document$\rangle$ pair. 
As shown in the upper side of Figure~\ref{fig:cda_example}, during training, models learn to predict higher relevance scores for clicked documents and lower scores for unclicked documents~\cite{ase,ricr,hqcn}. 
While this training paradigm is intuitive and effective, it neglects an important fact ---\textit{the relevance between the search context and candidate document is symmetric}.
Let us analyze the relevance of a positive pair: 
For the search context, the clicked document can fulfill the search intent more effectively than others. 
This has been considered in the existing training paradigm (as shown in the upper part of Figure~\ref{fig:cda_example}). 
In contrast, for the clicked document, the search context should also be the one that matches its content the most. 
This aspect is unfortunately missed by existing methods, resulting in insufficient learning.
To put it another way, existing methods are not able to teach the models that the relevance of a document could be different when the session context changes. 
Consider the following scenario: Suppose a user's current query is ``Artificial Intelligence'' and their previous search was for ``Machine Learning Algorithms''. In this situation, the user likely seeks information about AI algorithms or related topics. However, if the user's previous search was for ``Job Opportunities in Tech'', or if their current query changes to ``Online Courses on Programming'', the relevance ranking of the candidate documents should be different.
The problem is even more severe for the PLM-based methods, as they always constructed training sequences by fixed search context and different candidate documents.

To address this problem, we propose to augment the training data by considering search context alterations (as shown in the lower part of Figure~\ref{fig:cda_example}), \ie, fixing the clicked document and identifying possible alternations of the search context to construct more training pairs.
In general, a search context consists of three components: historical queries, corresponding clicked documents, and the current query. The decision to modify the current query is based on two key observations:
(1) The current query is the most effective information in the search context to understand the user's search intent. 
To support this, we conduct a preliminary experiment based on a well-known baseline {COCA}~\cite{coca}: To evaluate the impact of the current query and the historical query-document pairs on the performance of COCA, we proceed by removing both the current query and the last two query-document pairs from the session history, respectively. The performance is evaluated in terms of Mean Average Precision (MAP), Mean Reciprocal Rank (MRR), and Normalized Discounted Cumulative Gain (NDCG) at position $k$ (NDCG@$k$), where $k$ takes values from the set $\{3,10\}$.
The results shown in Table~\ref{tab:pre_exp} indicate that the absence of the current query has the most significant impact on ranking performance.
(2) Context-aware ranking models concentrate on representing the entire search behavior sequence, which may weaken its modeling of the current query. Enhancing the model's ability to capture more fine-grained information from the current query is important.

\begin{table}[t!]
    \centering
    \small
    \caption{Performance of COCA with different queries missing in the training data.
    Supposing $q_\text{c}$ denotes the current query. The query sequence that contains $n$ historical queries is $\{q_1, \cdots, q_{n-1}, q_{n}, q_{c}\}$.
    The corresponding clicked documents of the historical queries are $\{d_1, \cdots, d_{n-1}, d_{n}\}$. We remove the current query ($q_{c}$) and the last two query-document pairs ($q_{n}$ \& $d_{n}$, $q_{n-1}$ \& $d_{n-1}$), respectively. ``NDCG@$k$'' is referred to as ``N@$k$''.
    }
    \setlength{\tabcolsep}{1.0px}{
    \begin{tabular}{l|c|c|c|c}
    \toprule
         Metric & w/o. $q_{c}$ & {w/o. $q_{n}$ \& $d_{n}$} & w/o. $q_{n-1}$ \& $d_{n-1}$ & COCA \\
        \midrule
        MAP & 0.4751 \ \ -13.62\% & {0.5452 \ \ -0.88\%}  & 0.5465 \ \ -0.64\%  & \textbf{0.5500} \\
        
        MRR & 0.4860 \ \ -13.23\% & {0.5555 \ \ -0.83\%} & 0.5566 \ \ 
        -0.63\%  & \textbf{0.5601} \\
        
        N@3 & 0.4631 \ \ -15.46\% & {0.5416 \ \ -1.14\%}  & 0.5429 \ \ -0.90\%  & \textbf{0.5478} \\
        N@10 & 0.5450 \ \ -11.53\% & {0.6103 \ \ -0.93\%}  & 0.6120 \ \ -0.65\%  &  \textbf{0.6160} \\
    \bottomrule
    \end{tabular}}
    \label{tab:pre_exp}
\end{table}

More specifically, we propose a \textbf{Q}uery-oriented Data \textbf{A}ugmentation method for \textbf{S}ession \textbf{S}earch (\ours{}). 
We generate new training samples by altering the current query to complement real-world search logs and facilitate model learning. 
In the training process, the generated samples serve as \textbf{negative} samples, given that the original samples are directly observed in the search log.
Specifically, we consider altering the current query at two levels: 
(1) Term-level Modification. By changing (\ie,  masking, replacing, or adding) some terms within the current query, the model can learn the impact of subtle variations in the query.
(2) Query-level Replacement. 
We directly replace the current query with some queries mining from the search log. 
In this process, we also consider the difficulty of query modification, inspired by recent studies in dense retrieval~\cite{ambineg,KarpukhinOMLWEC20,XiongXLTLBAO21}, where a mixture of negative documents in different difficulties can make the training process more stable~\cite{dcl,jingtao}. 
In particular, all samples generated by randomly sampled queries from the search log are considered ``easy'' negative samples. 
Then, we replace the current query with its historical queries in the search context. 
The generated samples are treated as ``medium'' negative samples because the replaced query is close to the current query (they appear in the same session). 
Similarly, the queries augmented by term-level modification are also used as ``medium'' negative samples.
Finally, we mine some ambiguous queries of the current query by some heuristics. 
We use the generated samples as ``hard'' negative samples because the ambiguous queries are even closer to the current query than the historical ones. 
Through these strategies, we can generate negative sequences of varying difficulty with respect to the current queries.
Our experiments on two public search logs (AOL~\cite{aol} and Tiangong-ST~\cite{tiangong}) demonstrate that \ours{} significantly outperforms existing models, indicating the effectiveness of our proposed query-oriented data augmentation method.

In summary, the contributions of the paper are as follows:

(1) We identify the problem in current training paradigms, where the relevance from the perspective of the clicked document is overlooked.
We propose to generate query-oriented data for session search by altering the current query, thereby enriching search logs and enabling models to learn users' search patterns more comprehensively.
It is the first time that negative sampling is performed on the query side rather than the document side for session search. 

(2) We develop various methods to generate negative training samples with varying difficulty.
Different score margins are applied to identify their difficulty and coordinate these augmented pairs.

(3) We design a heuristic for mining ambiguous queries, ensuring their similarity to the current query by considering the ranking of the clicked document in other sessions. 
Experimental results validate that these queries are more informative than other mined queries for learning users' search intents.

\section{Related Work} \label{sec:RW}

\subsection{Data Augmentation for Ranking}

There are already some research works that designed various data augmentation strategies to facilitate information retrieval models~\cite{nugraha2019typographic,DBLP:conf/icml/0001I20,DBLP:conf/cvpr/He0WXG20,DBLP:conf/ictir/LiLX021,DBLP:conf/acl/YangJLGC20,DBLP:conf/acl/YangWJJY20,DBLP:conf/eccv/FuWPGEW20,DBLP:conf/acl/ChenKNGZOM20,DBLP:conf/coling/YaoYZCL20,DBLP:conf/iccv/0016Z00PC19}
Moreover, data augmentation techniques can be applied to generate additional training data for document ranking models~\cite{dataaug2019icde,dataaug2019aaai,dataaug2021cikm}.
Through the generated synthetic data, the model can learn from a more diverse set of examples, which can improve its ability to rank documents effectively. 
Data augmentation can help in addressing issues like data sparsity, overfitting, and generalization, leading to better performance in document ranking tasks.
For example, Li et al.~\cite{dataaug2019icde} proposed an attention-based sequence-to-sequence model for POI recommendation.
Specifically, they incorporated spatial and temporal information to augment the check-in datasets.
Subsequently, an encoder-decoder model is applied to learn the missing check-in.
Yu et al.~\cite{dataaug2019aaai} designed an informative data generation model to address the data imbalance problem in learning to rank.
Based on the adversarial autoencoder, they disentangled the relevance information from the latent representation and exploited query information to regularize the prior distribution.
Qiu et al.~\cite{dataaug2021cikm} proposed a Learning to Augment (LTA) method to resolve the data imbalance issue.
They proposed to generate informative data using a Gaussian Mixture Variational Autoencoder.
Furthermore, they applied a teacher model to learn how to optimize their generation policy based on reinforcement learning. 
In Named Entity Recognition (NER) task, there are also some works utilizing data augmentation techniques to train better models~\cite{DBLP:conf/sisap/BartoliniMPSV22,DBLP:journals/is/BartoliniMPSV23}. For example, COSINER~\cite{DBLP:conf/sisap/BartoliniMPSV22} replaced entity mentions with alternatives, considering available training data and the contexts in which entities commonly occur.

\subsection{Modeling Search Sessions}
Some early works have resorted to statistical methods to study contextual information of search sessions.
Shen et al.~\cite{Shen2005} employed context-sensitive retrieval-based algorithms that rely on statistical language models to effectively incorporate session context.
Bennett et al.~\cite{BennettWCDBBC12} demonstrated that a combination of historical behaviors and short-term behaviors can benefit the understanding of search intents in a statistical manner.
White et al.~\cite{White2013} mined data from similar search sessions conducted by other users to identify documents that would be highly relevant.
Van Gysel et al.~\cite{VanGysel2016} studied lexical query modeling in session search.
They pointed out that context-aware methods are more effective than traditional query terms re-weighting.
These traditional approaches have achieved great success.
However, restricted by their non-parametric and statistic-based nature, they are not able to model user behaviors thoroughly.

The advent of deep learning has led to the emergence of numerous neural context-aware ranking models in recent years.
Ahmad et al.~\cite{mmtensor} encoded sequential historical behaviors and candidate documents with RNNs and computed the ranking score based on the matching of their representations.
They~\cite{cars} complemented their work using attention mechanisms and jointly learning the ranking task and the query suggestion task.
HBA-Transformer ~\cite{hba} concatenated the session sequence and used the popular PLM BERT as the encoder.
They also designed a hierarchical behavior-aware module to capture interaction-based information.
Zuo et al.~\cite{hqcn} modeled multi-level historical query changes to obtain representations of sessions from multiple aspects.
RICR~\cite{ricr} integrated representation and interaction. 
They employed Recurrent Neural Networks (RNNs) to effectively capture session sequences. This modeling technique was then utilized to augment the word-level interaction between the current query and candidate documents.
COCA~\cite{coca} employed data augmentation and contrastive learning methods to pre-train an enhanced BERT encoder for effectively modeling session sequences.
HEXA~\cite{hexa} utilized heterogeneous graphs to capture information within a session and from other sessions. 
DCL~\cite{dcl} designed a curriculum learning framework that learns the matching between the session context and documents from easy to hard.
Since it is a learning framework rather than a specific model and it re-samples the negative documents, we will omit it in our comparisons.
ASE~\cite{ase} used a decoder and several generation tasks specifically designed for session search to enhance the ability of the encoder.


Our approach to data augmentation differs from that of {COCA}~\cite{coca} in a significant way. 
While Zhu et al. treated the augmented sequences as positive examples of contrastive learning, we consider our generated sequences as negative examples in pair-wise training. 
This distinction arises from our focus on altering the most crucial behavior, namely the current query, rather than making slight changes to the session context, as suggested by Zhu et al. 
They argue that these minor changes should not affect the sequence representation significantly.
In contrast, we focus on altering the most important behavior, \ie, the current query, and we believe our augmentation strategies (\eg, replacing it with a random query) should make the search intent change, even under the same session history.
Our objective is to complement the existing training paradigm by generating alternate behavior sequences from the query side, whereas COCA aims to pre-train the encoders through an entirely separate training stage.

\section{Proposed Model: \ours{}} \label{sec:method}
In this work, we propose to pair various search contexts with the same clicked document, allowing the model to learn their distinct relevance. 
We choose to alter the current query to generate new search contexts, thus our approach can be treated as a query-oriented data augmentation method. 
Our data augmentation strategies involve altering terms in the current query (masking, replacing, or adding), and replacing the entire query with other queries mined from the search log (random queries, historical queries, and ambiguous queries).
These mined/generated queries form negative training pairs with different difficulty levels, which can help stabilize the training process.

\begin{figure}[t!]
    \centering
    \includegraphics[width=.9\linewidth]{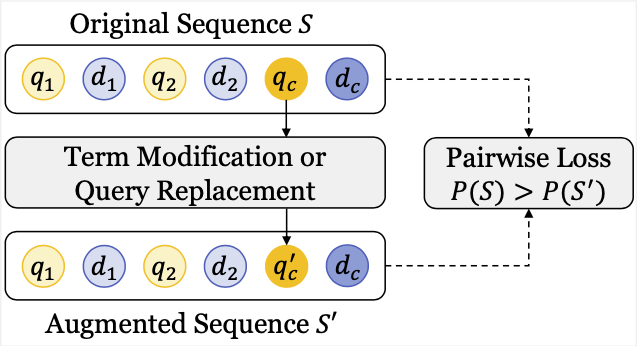}
    \caption{Illustration of \ours{}. The current query $q_\text{c}$ of the original user behavior sequence $S$ is altered to construct the augmented sequence $S'$. The clicked document $d_\text{c}$ is hypothesized to be more relevant to the original search context than the altered context (\ie, $P(S)>P(S')$).}
    \label{fig:model}
\end{figure}

\subsection{Important Notations}
Before introducing our model, we first explain some important notations of session search.
The target of this task is to model sequential user behaviors in a session to understand the user's search intent and rank the clicked documents as high as possible.
We denote the user's queries in the session history $H$ as $[q_1, q_2, \dots, q_n]$, and their corresponding clicked documents as $[d_1, d_2, \dots, d_n]$.\footnote{Following previous works~\cite{ase,coca}, we only use the first clicked document of each historical query.}
The current query is denoted as $q_\text{c}$, and its candidate document set is denoted as $D$. Furthermore, the clicked documents in $D$ are denoted as $d_\text{c}$, and the skipped documents are $d_\text{s}$ ($d_\text{c}\cup d_\text{s}=D, d_\text{c}\cap d_\text{s}=\varnothing$).

A context-aware document ranking model attempts to score $d \in D$ based on $H$ and $q_\text{c}$ as follows:
\begin{equation}
    P(d) = P(H, q_\text{c}, d). \label{eq:score}
\end{equation}
Existing training paradigm optimizes the model to rank $d_\text{c}\in d_\text{c}$ higher than $d_\text{s}\in d_\text{s}$, \ie, $P(H, q_\text{c}, d_\text{c}) > P(H, q_\text{c}, d_\text{s})$.
We enrich search logs with the data generated by altering the current query $q_\text{c}$ to $q_\text{c}'$ and let the model learn $P(H, q_\text{c}, d_\text{c}) > P(H, q_\text{c}', d_\text{c})$.

\subsection{Model Overview}
In this section, we will provide a concise overview of the structure of our model.
Our model is comprised of two stages:

(1) \textbf{Query-oriented Data Augmentation.} 
We aim to generate query-oriented data pairs to enrich the search log by altering the current query $q_\text{c}$.
Sequences constructed from generated/mined queries are considered negative sequences compared to the actually observed ones.
Specifically, as shown in Table~\ref{tab:variation}, we employ term-level modification and query-level replacement to generate various queries, which are then used to construct negative training pairs with different difficulties.
For term-level modification, we change (\ie, mask, replace, or add) some terms of $q_\text{c}$, which enable \ours{} to learn fine-grained matching signals.
For query-level replacement, we mine some queries from search logs (including random queries, historical queries, and ambiguous queries) and replace $q_\text{c}$ with these queries, helping our model extract search intent at a higher level.

(2) \textbf{Jointly Training on All Data Pairs.} 
As illustrated in Figure~\ref{fig:model}, with the generated and original data pairs ready, we use the pre-trained language model BERT to score all sequences and apply a pair-wise loss function to optimize the model. 
To identify training pairs of varying difficulty, we apply different score margins for them. 

\subsection{Query-oriented Data Augmentation Strategies} \label{sec:da_stra}
As shown in Figure~\ref{fig:model} and Table~\ref{tab:variation}, we alter $q_\text{c}$ at both the term and query levels. 
These modifications serve distinct purposes: the term-level modification introduces slight variations to the original query, which enhances the model's capability to capture fine-grained interactions.
On the other hand, query-level replacement directly changes the entire query, requiring the model to understand the query from a higher view.
\footnote{Since \ours{} strives to emphasize that changes in the current query have a significant influence on the search intent of the search sequence even under the same history, we only alter the queries that have historical queries.}

\begin{table}[t]
    \centering
    \small
    \caption{Examples of queries generated by different data augmentation strategies. 
    We take an actual session from the AOL search log as an example: $H$ is \{ ``racine county history'' ($q_1$), ``racine county wi home'' ($d_1$) \}, $q_\text{c}$ is ``burlington wisconsin'', and $d_\text{c}$ is ``burlington wi official website''.
    Texts in bold indicate unchanged terms.}
    \setlength{\tabcolsep}{3.5px}\begin{tabular}{lllr}
    \toprule
        Level & Type & Query & Difficulty \\
    \midrule
        - & Original & burlington wisconsin & - \\
    \midrule
        Term & Mask & \textbf{burlington} [term\_del] & Medium \\
        Term & Replace & \textbf{burlington} becker & Medium \\
        Term & Add & school \textbf{burlington wisconsin} & Medium \\
    \midrule
        Query & Random & laugh factory nyc & Easy \\
        Query & Historical & racine county history & Medium \\
        Query & Ambiguous &  \textbf{burlington} county jobs & Hard \\
    \midrule
        \multicolumn{4}{l}{Search session: } \\
        \multicolumn{4}{l}{$H$: [$q_1$: racine county history, $d_1$: racine county wi home]} \\
        \multicolumn{4}{l}{$q_\text{c}$: burlington wisconsin, \quad $d_\text{c}$: burlington wi official website} \\
    \bottomrule
    \end{tabular}
    \label{tab:variation}
\end{table}

\subsubsection{Term-level Modification} \label{sec:tlm}

Some existing works in Natural Language Processing (NLP) have used word-level augmentation to make the representations of sentences more robust~\cite{term1,term2,term3}.
A recent model for session search, COCA~\cite{coca}, also generates sequences for contrastive learning by masking terms within session sequence.
Inspired by these studies, we use the term-level modification to the current query to construct additional data.

Most queries in actual search logs primarily comprise keywords~\cite{ase}, namely, they contain fewer than three terms (approximately 72.5\% in AOL search log). 
Thus, subtle term-level modifications of $q_\text{c}$ may result in noticeable changes in search intent.
Supposing $q_\text{c} = \{w_1, \cdots, w_t\}$, we design three term-level modification strategies:

(1) \textbf{Term Masking.}
We randomly select an index $k$ ranging in $[0, t]$, and mask the word $w_k$ by replacing it with ``[term\_del]'' (similar to the [MASK] token in BERT) as:
\begin{align}
    q_\text{c}' = [w_1, \dots, w_{k-1}, {\rm [term\_del]}, w_{k+1}, \dots, w_t].
\end{align}
For example, $q_\text{c}'$ ``burlington [term\_del]'' is generated from $q_\text{c}$ ``burlington wisconsin'' by term masking.
Obviously, $q_\text{c}'$ lacks the important information ``wisconsin'', which makes the current search intent different, even with the same session history $H$. 
Thus, the generated sequence $[H, q_\text{c}']$ should be less relevant to the clicked document $d_\text{c}$.

(2) \textbf{Term Replacing.}
We first randomly select an index $k$ ranging in $[0, t]$.
Then, we randomly mine a term $w_r$ from the training data of search log and replace $w_k$ with $w_r$ as:
\begin{align}
q_\text{c}' = [w_1, \dots, w_{k-1}, w_r, w_{k+1}, \dots, w_t],
\end{align}
where $w_r$ should be different from $w_k$. 
Similar to term masking, this strategy may also change the search intent of the current query.

(3) \textbf{Term Adding.}
First, we randomly select an index $k$ ranging in $[0, t+1]$.
We then randomly mine a term $w_a$ from the search log and insert it at the position $k$ as:
\begin{align}
q_\text{c}' = [w_1, \dots, w_{k-1}, w_a, w_{k}, \dots, w_t].
\end{align}
The added word may introduce a more complex search intent to $q_\text{c}$.

\subsubsection{Query-level Replacement} \label{sec:qlr}

In addition to term-level augmentations, we alter the whole query from a higher view, \ie, mine some queries to directly replace $q_\text{c}$. 
Specifically, we mine three kinds of queries from the search log:

(1) \textbf{Random Query.}
The most straightforward way is to replace $q_\text{c}$ with a randomly mined query from the search log.
The sequences generated by this strategy may contain much noise, which can make our model more robust.

(2) \textbf{Historical Query.}
A user's search behaviors in the same session usually serve for a main search intent~\cite{ase, jones2008, Wang2013, coca, cars}, while queries in the same session can usually represent different subtle intents.
For example, as shown in Table~\ref{tab:variation}, $q_1$ and $q_\text{c}$ both try to find the websites of certain cities in Wisconsin.
However, $q_1$ tries to find the history of Racine county, whereas $q_\text{c}$ wants to learn about Burlington.
Thus, these two queries share some common intents but are still not identical, which makes $q_1$ a good replacement for $q_\text{c}$ to train the model. 
Compared to the random query, the historical queries are more similar to the current query, so the generated training pairs are more difficult.

\begin{figure}[t]
    \centering
    \includegraphics[width=.7\linewidth]{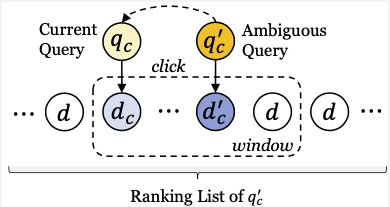}
    \caption{Illustration of mining the ambiguous queries.
    We first use a ranking model to obtain a ranking list of all documents for each query.
    Then a window of negative documents is sampled around each query's clicked document.
    If $d_\text{c}$ is in the window of a query $q_\text{c}'$, this query $q_\text{c}'$ is an ambiguous query of $q_\text{c}$.
    The closer $d_\text{c}$ is to the clicked document of $q_\text{c}'$ ($d_\text{c}'$), the more ambiguous $q_\text{c}'$ is to $q_\text{c}$.
    }
    \label{fig:ambi}
\end{figure}

(3) \textbf{Ambiguous Query.}
Inspired by a recent study on mining ambiguous documents for dense retrieval~\cite{ambineg}, we propose to mine some ambiguous queries to replace the current query.
Building on this insight, our objective is to mine negative queries that strike a balance between being too challenging (possibly false negatives) and too easy (uninformative) to replace the current query.
Intuitively, if the clicked document $d_\text{c}$ under the current query $q_\text{c}$ is ranked around the clicked document $d_\text{c}'$ under another query $q_\text{c}'$ by a ranking model, we can treat $q_\text{c}'$ as an ambiguous query of $q_\text{c}$.
This is because their respective clicked documents are very similar, indicating a potential overlap in search intent.
As shown in Figure~\ref{fig:ambi}, we first rank all documents for each query in the search log so that each query has a complete document ranking list.
Subsequently, for each clicked document associated with a query, we sample a window of documents using the clicked document as the center. 
The documents within the window are highly relevant to the clicked document.
Finally, in order to mine ambiguous queries for $q_\text{c}$, we choose the queries whose document window contains $d_\text{c}$.
Following~\cite{dcl}, to get the ranking list of each query with good quality and efficiency, we train a dense retriever based on the BERT~\cite{bert} representation of queries and documents with the dot-product as the relevance score, namely:
\begin{align}
    P(q,d) = {\operatorname{BERT}}(q)_{\rm [CLS]} \cdot {\operatorname{BERT}}(d)_{\rm [CLS]}.
\end{align}
We use in-batch negatives for training and FAISS~\cite{faiss} to achieve fast retrieval. 

As yet, we have generated $q_\text{c}'$ with various strategies. 
Then, we use $q_\text{c}'$ to replace $q_\text{c}$ and generate new search behavior sequences. 
Finally, they are combined with the clicked document $d_\text{c}$ to form new negative training pairs.
Based on the similarity between $q_\text{c}'$ and $q_\text{c}$, we can categorize the difficulty of the generated negative pairs into three levels:
(1) The random queries obtained in query-level replacement are used to form the least difficult negative pairs.
This is intuitive since these queries are randomly sampled from other search sessions, which may have totally different search intents.
(2) The historical queries and those generated by term-level modifications are used to construct negative pairs of medium difficulty.
These altered or replaced queries may share some common intents with $q_\text{c}$ yet are still different.
Thus we consider them both as ``medium negative'' pairs for simplicity.
(3) The ambiguous queries are used to form hard negative pairs. 
According to our heuristic of mining ambiguous queries, their search intent may be very close to the current query, so the generated negative pairs are the most difficult to distinguish. 

\subsection{Scoring and Training}
\label{sec:train}
\subsubsection{Sequence Scoring with BERT}
Pre-trained language models, such as BERT~\cite{bert}, are widely used in the task of session search~\cite{hba,coca,ase}.
We use BERT as the backbone model for a fair comparison with existing methods.
Following~\cite{coca}, we first use some special tokens to concatenate behaviors in $H$: $I_H = q_1 {\rm [EOS]} d_1 {\rm  [EOS]} \cdots q_n{ \rm [EOS]} d_n $, where the ``${\rm [EOS]}$'' token indicates the end of a query/document. 
Then, we append the current query $q_\text{c}$ and the candidate document $d$ to $I_H$: $I_S = {\rm [CLS]} I_H {\rm [EOS]} q_\text{c} {\rm [EOS]} {\rm [SEP]} d {\rm [EOS]} {\rm [SEP]}$, where ${\rm [SEP]}$ is the separator to identify the candidate document $d$, ${\rm [CLS]}$ is used for computing the sequence representation.
Then we use BERT to compute the ranking score of the sequence [$H, q_\text{c}, d$]: 
\begin{equation}
    P(H, q_\text{c}, d) = {\rm MLP} ({\rm BERT_{[CLS]}}(I_S)),
\end{equation}
where ${\rm MLP}$ is a multi-layer perceptron used as a classifier to compute the ranking score.

\subsubsection{Jointly Optimizing All Training Pairs} \label{subsec:training}
Taking the training of a query $q_\text{c}$ as an example, the (search context, clicked document) pairs that are actually observed in the search log are used to construct the positive sequences, \ie, $S_\text{p} = [H, q_\text{c}, d_\text{c}]$. 
The training objective is to predict the ranking score of the positive sequence $S_\text{p}$ to be higher than that of the negative sequence $S_\text{n}$.
A standard pair-wise ranking function is often employed to optimize the model:
\begin{equation}
    \mathcal{L}(S_\text{p}, S_\text{n}) = 
    {\rm max} \left( 0, m - P(S_\text{p}) + P(S_\text{n}) \right), 
\end{equation}
where $m$ is a hyperparameter, representing the minimum acceptable score margin between $S_\text{p}$ and $S_\text{n}$.

In the original datasets, the skipped documents $d_\text{s}$ are used to construct negative training pairs, \ie, $S_\text{n}=[H, q_\text{c}, d_\text{s}]$. They are also valuable for learning document ranking, so we keep them for training as follows:
\begin{equation}
    \mathcal{L}_{\text{op}}(q_\text{c}) = \sum_{(d_\text{c},d_\text{s}) \in D}^{} 
    {\rm max} \left( 0, m_{\text{op}} - P(S_\text{p}) + P(H, q_\text{c}, d_\text{s}) \right), \notag
\end{equation}
where $m_{\text{op}}$ is a margin hyperparameter, which is usually set to 1.0 for binary classification problems.

We also apply our data augmentation strategies to generate negative sequences, in which the current query $q_\text{c}$ is altered to $q_\text{c}'$. 
These negative sequences can be represented as $S_\text{n}=[H, q_\text{c}', d_\text{c}]$. The training objective is defined as:
\begin{equation}
    \mathcal{L}_{\text{cp}}(q_\text{c}) = \sum_{d_\text{c} \in D}^{} \sum_{q_\text{c}'}^{} 
    {\rm max} \left( 0, m_{\text{cp}} - P(S_\text{p}) + P(H, q_\text{c}', d_\text{c}) \right), \notag
\end{equation}
where $m_{\text{cp}}$ is the margin hyperparameter for constructed pairs.

To identify sequences generated by different strategies, we apply different margins during the training process.
The margin can control the distance between $S_\text{p}$ and $S_\text{n}$, \ie, the smaller the margin is, the more relevant $q_\text{c}'$ is to $d_\text{c}$.
Following the discussion of difficulty before, $m_{\text{cp}}$ have three levels: $m_{\text{rq}}$ (random queries) $\textgreater$ $m_{\text{th}}$ (term-level modification and historical queries) $\textgreater$ $m_{\text{aq}}$ (ambiguous queries).
The intuition here is that the higher the ranking of $d_\text{c}$, the closer $q_\text{c}'$ and $d_\text{c}$, \ie, the smaller the margin.
The influence of the score margins is studied in Section~\ref{margin_exp}.
Specifically, for $m_{\text{aq}}$, we slightly tune each ambiguous query's margin based on the position of $d_\text{c}$ in that ambiguous query's ranking list: $m_{\text{aq}} = ({\rm pos}(d_\text{c})/w_{\text{size}})*2*\overline{m}_{\text{aq}}$. 
$\overline{m}_{\text{aq}}$ is the average margin of ambiguous queries, and $w_{\text{size}}$ is the size of the window of negative documents.

In summary, for a query $q_\text{c}$, we train the original data pairs and the augmented data pairs jointly:
\begin{equation}
    \mathcal{L}(q_\text{c}) = \mathcal{L}_{\text{cp}}(q_\text{c}) + \mathcal{L}_{\text{op}}(q_\text{c}). 
\end{equation}

\section{Experiments} \label{sec:setting}

\subsection{Datasets and Evaluation}

\subsubsection{Datasets}

\begin{table}[t!]
    \centering
    \small
    \caption{Statistics of two search logs.}
    \begin{tabular}{lrrrrrr}
    \toprule
        \textbf{AOL} & {\textbf{Training}} & {\textbf{Validation}} & {\textbf{Testing}} \\
    \midrule
        \# session & 219,748 & 34,090 & 29,369 \\
        \# query  & 566,967 & 88,021 & 76,159 \\
        avg. session length & 2.58 & 2.58 & 2.59 \\
        avg. query length & 2.86 & 2.85 & 2.9 \\
        avg. document length & 7.27 & 7.29 & 7.08 \\ 
        \# candidate per query & 5 & 5 & 50 \\
        avg. \# click per query & 1.08 & 1.08 & 1.11 \\
    \midrule
        \textbf{Tiangong-ST} & {\textbf{Training}} & {\textbf{Validation}}  & {\textbf{Testing}} \\
    \midrule
        \# session & 143,155 & 2,000 & {2,000}\\
        \# query  & 344,806 & 5,026 &  {6,420}\\
        avg. session length & 2.41 & 2.51 & {3.21}\\
        avg. query length & 2.89 & 1.83 & {3.46}\\
        avg. document length & 8.25 & 6.99 & {9.18}\\
       \# candidate per query & 10 & 10 & {10}\\
        avg. \# click per query & 0.94 & 0.53 & {3.65}\\
    \bottomrule
    \end{tabular}
    \label{tab:dataset}
\end{table}

Following existing works~\cite{ricr,ase,coca,hqcn}, we use two large-scale public search logs AOL~\cite{aol} and Tiangong-ST~\cite{tiangong} to compare the performance of \ours{} and baselines.
In accordance with previous research, we opt for utilizing only the document title to ensure efficiency. The statistical data derived from the analysis of these two search logs is illustrated in Table~\ref{tab:dataset}.

To use the AOL search log, we utilize the dataset provided by the authors of CARS~\cite{cars}. It is important to note that every query in this log has a minimum of one click that meets the user's satisfaction criteria. In both the training and validation sets, there are five candidate documents for each query, while the testing set contains 50 candidate documents.

The Tiangong-ST search log was obtained from a Chinese commercial search engine. 
It comprises user sessions spanning a duration of 18 days, including their top 10 search results and click information. 
However, certain queries and documents in the Tiangong-ST dataset are incomplete, and to address this, we have used the placeholders ``[empty\_q]'' and ``[empty\_d]'' to indicate missing content.
To create our test set, we have selected 2,000 sessions from the entire dataset. These sessions were chosen based on the condition that their last query possesses human relevance labels. For the remaining sessions, we have allocated the last 2,000 sessions as the validation set, while the remaining sessions form the training set.
It is important to note that only the last query in each session of the test set has an annotated relevance score, ranging from 0 to 4. This approach aligns with the methodology employed in recent studies~\cite{ase,dcl} as well as the original paper introducing this dataset~\cite{tiangong}. Consequently, during testing, we will exclusively utilize queries that possess relevance labels.
For more comprehensive information regarding this dataset, please refer to the original paper by Tiangong~\cite{tiangong}.




\begin{table*}[t!]
    \centering
    \caption{Overall results on two search logs.
    ``$\dag$'' denotes the result performs significantly worse than our model in paired t-test with $p$-value $<$ 0.01 and ``$\ddag$'' denotes a $p$-value $<$ 0.05 level.
    We highlight the the best performance in bold and the second-best one underlined. }
    \setlength{\tabcolsep}{4.0px}{\begin{tabular}{lcccccccccccccc}
    \toprule
        Dataset & Metric & {BM25} & {ARC-I} & {ARC-II} & {Duet} & {KNRM} & {CARS} & {HBA} & {RICR} & {HQCN} & BERT & {COCA} & {ASE} & {QASS} \\
        \midrule
        \multirow{6}{*}{AOL} & MAP & 0.2200$^\dag$ & 0.3361$^\dag$ & 0.3834$^\dag$ &  0.4008$^\dag$ & 0.4038$^\dag$ & 0.4297$^\dag$ & {0.5281}$^\dag$ & 0.5338$^\dag$ & 0.5448$^\dag$ & 0.5471$^\dag$ & {0.5500}$^\dag$ & \underline{0.5650}$^\dag$ & \textbf{0.5750}   \\
        
        & MRR & 0.2271$^\dag$ & 0.3475$^\dag$ & 0.3951$^\dag$ & 0.4111$^\dag$ & 0.4133$^\dag$ & 0.4408$^\dag$ & {0.5384}$^\dag$ & 0.5450$^\dag$ & 0.5549$^\dag$ & 0.5572$^\dag$ & {0.5601}$^\dag$ & \underline{0.5752}$^\dag$ & \textbf{0.5850}  \\
        
        & N@1 & 0.1195$^\dag$ & 0.1988$^\dag$ & 0.2428$^\dag$ & 0.2492$^\dag$ & 0.2397$^\dag$ & 0.2816$^\dag$ & {0.3773}$^\dag$ & 0.3894$^\dag$ & 0.3990$^\dag$ & 0.3990$^\dag$ & {0.4024}$^\dag$ & \underline{0.4144}$^\dag$ & \textbf{0.4266}  \\
        
        & N@3 & 0.1862$^\dag$ & 0.3108$^\dag$ & 0.3564$^\dag$ & 0.3822$^\dag$ & 0.3868$^\dag$ & 0.4117$^\dag$ & {0.5241}$^\dag$ & 0.5267$^\dag$ & 0.5441$^\dag$ & 0.5440$^\dag$ & {0.5478}$^\dag$ & \underline{0.5682}$^\dag$ & \textbf{0.5789}  \\
        
        & N@5 & 0.2136$^\dag$ & 0.3489$^\dag$ & 0.4026$^\dag$ & 0.4246$^\dag$ & 0.4322$^\dag$ & 0.4542$^\dag$ & {0.5624}$^\dag$ & 0.5648$^\dag$ & 0.5783$^\dag$ & 0.5818$^\dag$ & {0.5849}$^\dag$ & \underline{0.6007}$^\dag$ & \textbf{0.6104}  \\
        
        & N@10 & 0.2481$^\dag$ & 0.3953$^\dag$ & 0.4486$^\dag$ & 0.4675$^\dag$ & 0.4761$^\dag$ & 0.4971$^\dag$ & {0.5951}$^\dag$ & 0.5971$^\dag$ & 0.6070$^\dag$ & 0.6123$^\dag$ & {0.6160}$^\dag$ & \underline{0.6283}$^\dag$ & \textbf{0.6373}  \\  

        \midrule
        \multirow{4}{*}{Tiangong-ST}
        
        & N@1 & 0.6029$^\dag$ & 0.7088$^\dag$ & 0.7131$^\dag$ & 0.7577$^\dag$ & 0.7560$^\dag$  & 0.7385$^\dag$ & {0.7612}$^\dag$ & 0.7670$^\ddag$ & 0.7739$^\ddag$ & 0.7488$^\dag$ & {0.7769} & \underline{0.7884} & \textbf{0.7955} \\
        
        & N@3 & 0.6646$^\dag$ & 0.7087$^\dag$ & 0.7237$^\dag$ & 0.7354$^\dag$ & 0.7457$^\dag$ & 0.7386$^\dag$ & {0.7518}$^\dag$ & 0.7636$^\ddag$ & 0.7682 & 0.7541$^\dag$ & {0.7576}$^\ddag$ & \underline{0.7727} & \textbf{0.7742} \\
        
        & N@5 & 0.7072$^\dag$ & 0.7317$^\dag$ & 0.7379$^\dag$ & 0.7548$^\dag$ & 0.7716$^\dag$ & 0.7512$^\dag$ & {0.7639}$^\dag$ & 0.7740$^\ddag$ & 0.7783$^\ddag$ & 0.7651$^\dag$ & {0.7703}$^\dag$ & \underline{0.7839} & \textbf{0.7861} \\
        
        & N@10 & 0.8541$^\dag$ & 0.8691$^\dag$ & 0.8732$^\dag$ & 0.8829$^\ddag$ & 0.8894$^\ddag$ & 0.8837$^\ddag$ & {0.8896}$^\ddag$ & 0.8934$^\ddag$ & 0.8976 & 0.8890$^\ddag$ & {0.8932}$^\ddag$ & \underline{0.8996} & \textbf{0.9010} \\ 
    \bottomrule
    \end{tabular}}
    \label{tab:result}
\end{table*}

\subsubsection{Evaluation}

We evaluate the performance of \ours{} and baseline models using three metrics: Mean Average Precision (MAP), Mean Reciprocal Rank (MRR), and Normalized Discounted Cumulative Gain (NDCG) at position $k$ (NDCG@$k$) , where $k$ takes values from the set $\{1,3,5,10\}$. These metrics provide a comprehensive assessment of the effectiveness of the models:
\begin{align*}
{\rm{MAP}} &= \frac{1}{N} \sum_{i=1}^{N} \frac{1}{d_i^{\text{c}}} \sum_{j=1}^{d_i^{\text{c}}} \frac{j}{p_i^j}, \\
{\rm{MRR}} &= \frac{1}{N} \sum_{i=1}^{N} \frac{1}{p_i^1}, \\
\rm{DCG} (\sigma) &= \frac{1}{N} \sum_{i=1}^{N} \sum_{j=1}^{d_i^{\text{c}}} \frac{1}{{\rm log} (1+p_i^j)}, \\
{\rm{NDCG}} &= \frac{\rm{DCG} (\sigma_{\text{actual}})} {\rm{DCG} (\sigma_{\text{optimal}})} , 
\end{align*}
where $N$ represents the total number of queries in the evaluation set, $d_i^{\text{c}}$ denotes the number of user clicks that occurred during the $i$-th query, $p_i^j$ represents the position of the $j$-th click within the ranking list of the query $i$, and $\sigma$ is the ranking list.

Note that the Tiangong-ST dataset's relevance labels are human-annotated, hence the evaluation of MAP and MRR may not be accurate. 
We focus on NDCG@$k$ measures as suggested by the latest works~\cite{ase,hexa} and the original authors of this dataset~\cite{tiangong}.
Specifically, we use the tool provided by TREC to compute these metrics (trec\_eval~\cite{pytrec}).

\subsection{Baseline Models} \label{subsec:baselines}

Following~\cite{ricr,ase,coca,hqcn}, we evaluate \ours{} against two types of baselines:

(1) \textbf{Ad-hoc ranking models.} These models consider only the information of the current query $q_\text{c}$ and the candidate document $d$, \ie, neglect the session history.
\begin{itemize}
    \item {BM25}~\cite{bm25} BM25 is a traditional ranking algorithm that calculates the relevance between the current query $q_\text{c}$ and $d$ using probability.
    \item {ARC-I}~\cite{arci} utilizes convolutional neural networks (CNNs) to model both $q_\text{c}$ and $d$. 
    Then it computes these representations' semantic similarity as the relevance score.
    \item {ARC-II}~\cite{arci} uses 2D-CNNs to obtain the fine-grained interaction-based information of $q_\text{c}$ and $d$.
    \item {KNRM}~\cite{knrm} first construct a word-level interaction matrix of $q_\text{c}$ and $d$. Then it computes the relevance score based on soft matching signals by kernel pooling.
    \item {Duet}~\cite{duet} effectively evaluates the score of $d_\text{c}$ by incorporating a combination of interaction-based and representation-based features.
\end{itemize}

(2) \textbf{Context-aware ranking models}. These models utilize the session history to understand the search intent.
\begin{itemize}
    \item {CARS}~\cite{cars} jointly optimizes ranking and query suggestions. 
For ranking, it utilizes a Recurrent Neural Network (RNN) and attention to model sequential session behaviors.
\item {HBA-Transformers}~\cite{hba} uses BERT as the encoder of the session sequence.
In addition, It proposes a hierarchical behavior-aware attention module that is applied to BERT in order to extract detailed interaction-based information.
\item {HQCN}~\cite{hqcn} attempts to mine information from multi-granularity historical query change and utilize a query change classification task to help the ranking task.
\item {RICR}~\cite{ricr} utilizes Recurrent Neural Networks (RNNs) to encode the historical session information, followed by leveraging this encoded representation to improve the word-level performance of $q_\text{c}$ and $d$.
\item {BERT}~\cite{bert} is a pre-trained language model which is widely used in IR tasks.
\item {COCA}~\cite{coca} designs some data augmentation strategies to generate data for contrastive learning.
Additionally, it employs pre-training of BERT to improve the representation of the session sequence.
\item {ASE}~\cite{ase} uses a decoder and three generation tasks to enhance the encoding of the session sequence.
These generation tasks are designed specifically for session search.
\end{itemize}


\subsection{Model Settings} \label{subsec:implementation}
We use BERT provided by Huggingface as \ours{}'s backbone.\footnote{\url{https://huggingface.co/bert-base-uncased}}
We use AdamW~\cite{adamw} as the optimizer, and the training batch size is set as 588.
We train our model for three epochs and set the learning rate as 4e-5 with linear decayed.
For Tiangong-ST, we train data pairs of term-level modification after training other pairs because changes in Chinese characters may influence the training stability.

For each strategy of our query-oriented data augmentation, there are two kinds of hyperparameters: the number of generated sequences and the score margins.
(1) For term-level modification, we generate one sequence for each term-level strategy, \ie, three sequences in total.
For query-level replacement, we mine three random queries for AOL (eight for Tiangong), all historical queries in the session, and the four most ambiguous queries for AOL (five for Tiangong).
(2) We set $m_{\text{rq}}$ as 1.0, $m_{\text{th}}$ as 0.5, and $\overline{m}_{\text{aq}}$ as 0.2.
$w_{\text{size}}$ is set as 50.
More details of the implementation of \ours{} can be found in our anonymous repository, which is only for review.\footnote{\url{https://anonymous.4open.science/r/QASS-EFFF}}

\section{Experimental Results and Analysis} \label{sec:result}

\subsection{Overall Results} \label{subsec:overall result}

The experimental results on two search logs are presented in Table~\ref{tab:result}. It is observed that ad-hoc ranking models generally perform worse than models with contexts, highlighting the significance of incorporating session context in the modeling process. Additionally, the following observations can be made:

(1) \textbf{Our model \ours{} outperforms all other baselines.} 
It outperforms ASE, a strong baseline that enhances BART using multiple generative tasks, by approximately 2.94\% in terms of NDCG@1 on the AOL search log. 
This significant improvement showcases the capability of our generated data to enhance the existing training paradigm and provide valuable insights into user search patterns. 
Furthermore, it is worth noting that the improvements achieved by \ours{} on the AOL search log are more significant compared to those on the Tiangong-ST set.
This intriguing phenomenon was also noticed in previous works~\cite{ase,coca}.
We believe the possible reasons are: 
(i) \ours{} are trained on click-based search logs rather than relevance-based.
Our query-oriented data augmentation is also conducted on sequences of clicked documents, which makes \ours{} naturally perform better on predicting click behaviors than relevance scores.
(ii) The initial score on Tiangong-ST is already high. According to statistical analysis of the test set, more than 77.4\% of the candidate documents have relevance ratings greater than 1, meaning they are identified as relevant (Tiangong, citation). Even the basic neural model Duet achieves an impressive NDCG@10 score of 0.8829 on this dataset. Therefore, it becomes more challenging for our model, \ours{}, to show significant improvements on this particular dataset.

(2) \textbf{PLM-based models generally perform better than others.}
For example, the BERT-based models (COCA and \ours{}) and the BART-based model BART outperform RNN-based multi-task model CARS by over 20\% in terms of all metrics on the AOL search log.
The PLM-based models perform better than CARS even without the auxiliary task of query suggestion, which demonstrates the effectiveness of PLMs in modeling session context.

\subsection{Ablation Studies} \label{subsec:ablation}

To study the effectiveness of our data augmentation strategies, we conducted several ablation studies on AOL as follows:

$\bullet$ \textbf{\ours{} w/o. TM.} is \ours{} without the strategy of Term-level Modification (TM).

$\bullet$ \textbf{\ours{} w/o. RQ.} is \ours{} without the mined Random Queries (RQ) of query-level replacement, \ie, easy negatives.

$\bullet$ \textbf{\ours{} w/o. HQ.} is \ours{} without the Historical Queries (HQ) of query-level replacement.

$\bullet$ \textbf{\ours{} w/o. AQ.} is \ours{} without the Ambiguous Queries (AQ), \ie, hard negatives.

Due to the space limitation, we only present the performance of MAP, NDCG\\@3, and NDCG@5.
From the results in Table~\ref{tab:ablation}, we can find that all ablated models perform worse than \ours{}, which demonstrates the effectiveness of our data augmentation strategies.
Moreover, we can see:


\begin{table}[t!]
    \centering
    \caption{Performances of ablated models on AOL search log.}
    \setlength{\tabcolsep}{3.5px}{\begin{tabular}{l|cc|cc|cc}
    \toprule
         Metric & \multicolumn{2}{c}{MAP} & \multicolumn{2}{c}{NDCG@1} & \multicolumn{2}{c}{NDCG@3}  \\
        \midrule
        \ours{} (Full) & \textbf{0.5750} & - & \textbf{0.4266} & - & \textbf{0.5789} & - \\
        \quad w/o. TM & 0.5700 & -0.87\% & 0.4211 & -1.29\% & 0.5736 & -0.92\%  \\
        \quad w/o. RQ & 0.5706 & -0.77\% & 0.4217 & -1.15\%  & 0.5739 & -0.86\% \\
        \quad w/o. HQ & 0.5718 & -0.56\% & 0.4248 & -0.42\% & 0.5779 & -0.17\% \\
        \quad w/o. AQ & 0.5699 & -0.89\% & 0.4205 & -1.43\%  & 0.5735 & -0.93 \%  \\
    \bottomrule
    \end{tabular}}
    \label{tab:ablation}
\end{table}

\textbf{(1) Term-level modification can help \ours{} learn subtle modeling knowledge.}
We propose to change some words of $q_\text{c}$ to construct supplemental data pairs.
By this means, we try to help our model learn that subtle variations over the original query can result in changes in search intent.
This can enhance the model’s capability of capturing fine-grained information from the query.
After abandoning the data pairs generated by this strategy, \ours{}'s performance decreases by about 0.71\% in terms of NDCG@1.
This indicates our term-level modification strategy can help training.

\textbf{(2) Generating data pairs with random queries can make our model more robust.}
We randomly mine some queries from the search log to replace $q_\text{c}$.
These random queries may contain much noise, and the data pairs constructed by them can make our model more robust.
Specifically, removing these data makes the performance of \ours{} drop about 0.40\% in terms of NDCG@1.

\textbf{(3) Ambiguous queries are more informative than other queries.}
We attempt to mine some ambiguous queries to replace $q_\text{c}$ by tracking the ranking position of $d_\text{c}$ in other sessions.
Discarding these data results in the greatest decline of \ours{}.
This demonstrates that the data generated by ambiguous queries are the most informative ones.

\subsection{Influence of Ambiguous Query Sampling Strategies} \label{sec:ambi_exp}



As illustrated in Section~\ref{sec:qlr}, we mine ambiguous queries $q_c'$ from the search log by tracking the ranking position of $d_c$ in other sessions.
We sample a window of negative documents ($w_n$) around the clicked document of $q_c'$ (\ie, $d_c'$), and treat the proximity between $d_c$ and $d_c'$ as the ambiguity of $q_c'$.
We believe they are more informative than other generated queries, as demonstrated in the previous section.
In this section, we will further study the effectiveness of ambiguous queries sampled by different strategies from $w_n$.

We conduct our experiments by sampling ambiguous queries with different settings.
We treat the position of $d_c$ in $w_n$ as the difficulty of distinguishing $q_c'$ and $q_c$, \ie, the higher $d_c$ in $w_n$, the harder $q_c'$.
We first test different queries whose corresponding clicked documents $d_c'$ are in $w_n$: 
(1) ``Low queries'' are queries where $d_c$ is ranked low in their window, that is, $d_c$ is less relevant to $q_c'$ than $d_c'$ according to dot product by BERT.
(2) ``Medium queries'' are queries where $d_c$ is ranked in the medium part of $w_n$, \ie, around $d_c'$. 
These queries' $d_c'$ are closer to $d_c$ than low/high ranking ones, thus they are more ambiguous.
(3) ``High queries'' are queries where $d_c$ is ranked high in $w_n$, \ie, more relevant to $q_c'$.
We also test different retrieval models (BERT or BM25) for calculating query-document relevance for all queries.
We apply different strategies to sample the same number of ambiguous queries from $w_n$ and show the results in Table~\ref{tab:ambi_exp}.
From the results, we can have the following findings:

\begin{table}[tbp]
    \centering
    \caption{Performances of ambiguous queries sampled by different strategies from the negative document window (with different ranking positions of their clicked documents or different representation models) on AOL.
    The performance of ambiguous queries used in \ours{} is in bold.
    } 
    \begin{tabular}{lcccc}
    \toprule
        Strategies & MAP & MRR & NDCG@3 & NDCG@10 \\ 
        \midrule
        Low + BERT & {0.5740} & {0.5844} & {0.5776} & {0.6370} \\
        High + BERT & {0.5710} & {0.5812} & {0.5740}  & {0.6347} \\
        Medium + BERT &  \textbf{0.5750} &  \textbf{0.5850} &  \textbf{0.5789} &  \textbf{0.6373} \\
        Medium + BM25 & {0.5739} & {0.5840}  & {0.5773}  & {0.6372}  \\
    \bottomrule
    \end{tabular}
    \label{tab:ambi_exp}
\end{table}

\textbf{(1) Medium-ranking (Ambiguous) queries are more informative.}
It is clear to see that data pairs constructed from medium-ranking (ambiguous) queries perform better than both low-ranking and high-ranking queries.
We consider the reason is that the likelihood of false negatives resulting from the negative pairs created by ambiguous queries is reduced.
These negatives are neither too difficult (perhaps false negatives) nor too simple (uninformative).

\textbf{(2) The dense model BERT outperforms BM25 in terms of getting a high-quality ranking list.}
The results show that using BERT as the dense retriever when generating ranking lists performs better than the sparse retriever BM25.
The speculation is that we try to mine ambiguous queries at the semantic level rather than the term level (which we have already done in term-level modification), and the dense model BERT can better represent semantic-level information.

\begin{figure*}[tbp]
\centering
\includegraphics[width=\linewidth]{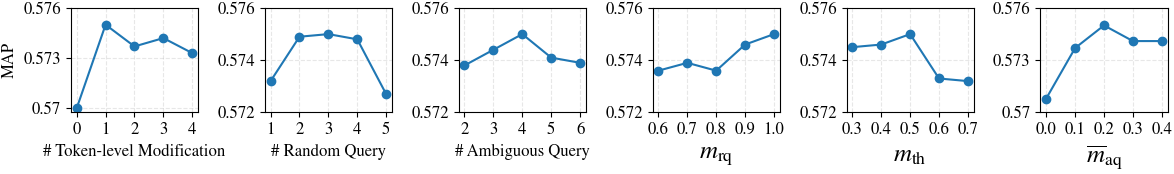}
\caption{Influence of the number of generated data pairs and the score margins.}
\label{fig:gen_num} 
\end{figure*}

\begin{figure}[tbp]
\centering
\includegraphics[width=\linewidth]{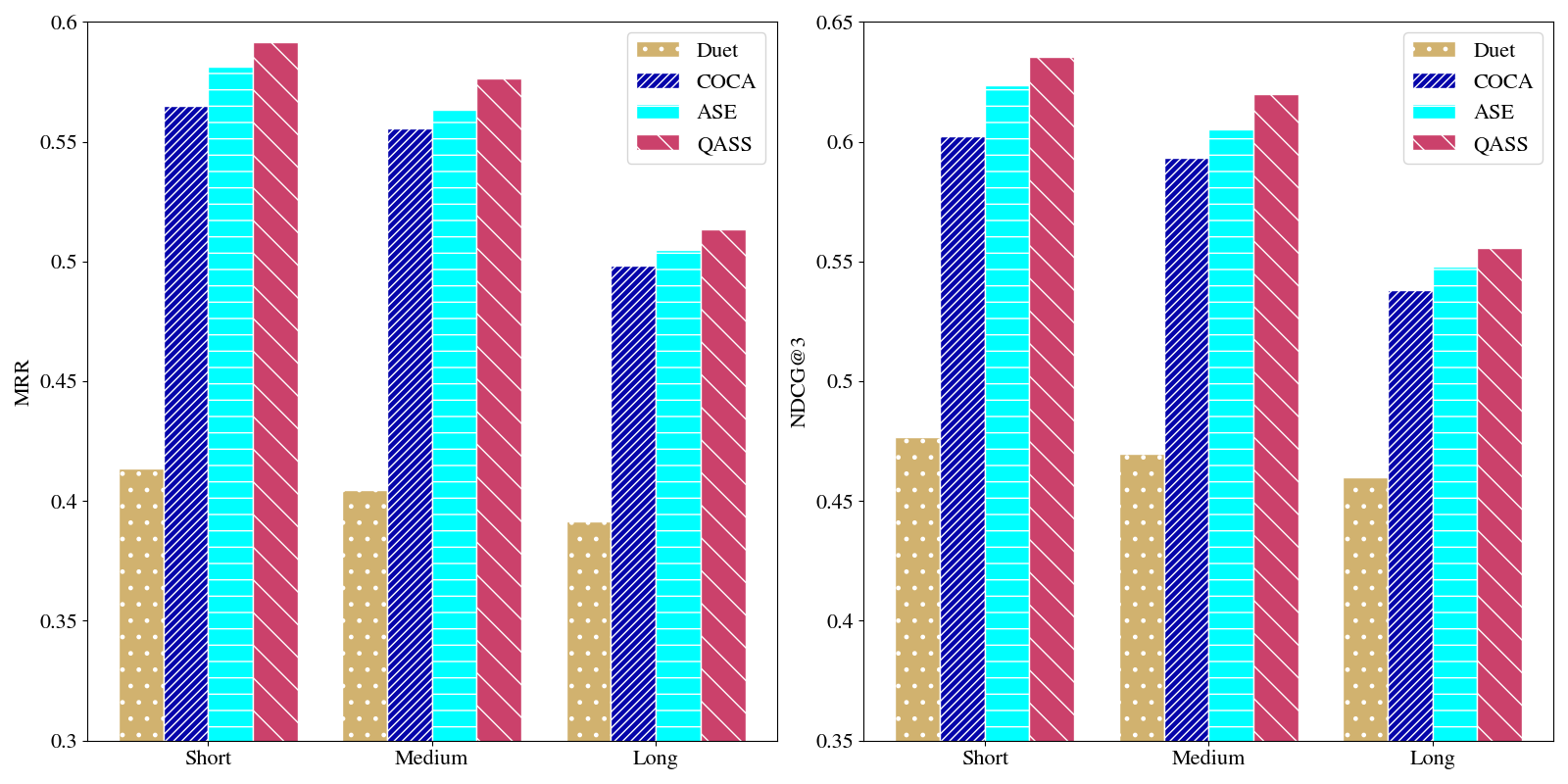}
\caption{Performances on different lengths of sessions on AOL search log.}
\label{fig:len} 
\end{figure}

\begin{figure}[tbp]
\centering
\includegraphics[width=\linewidth]{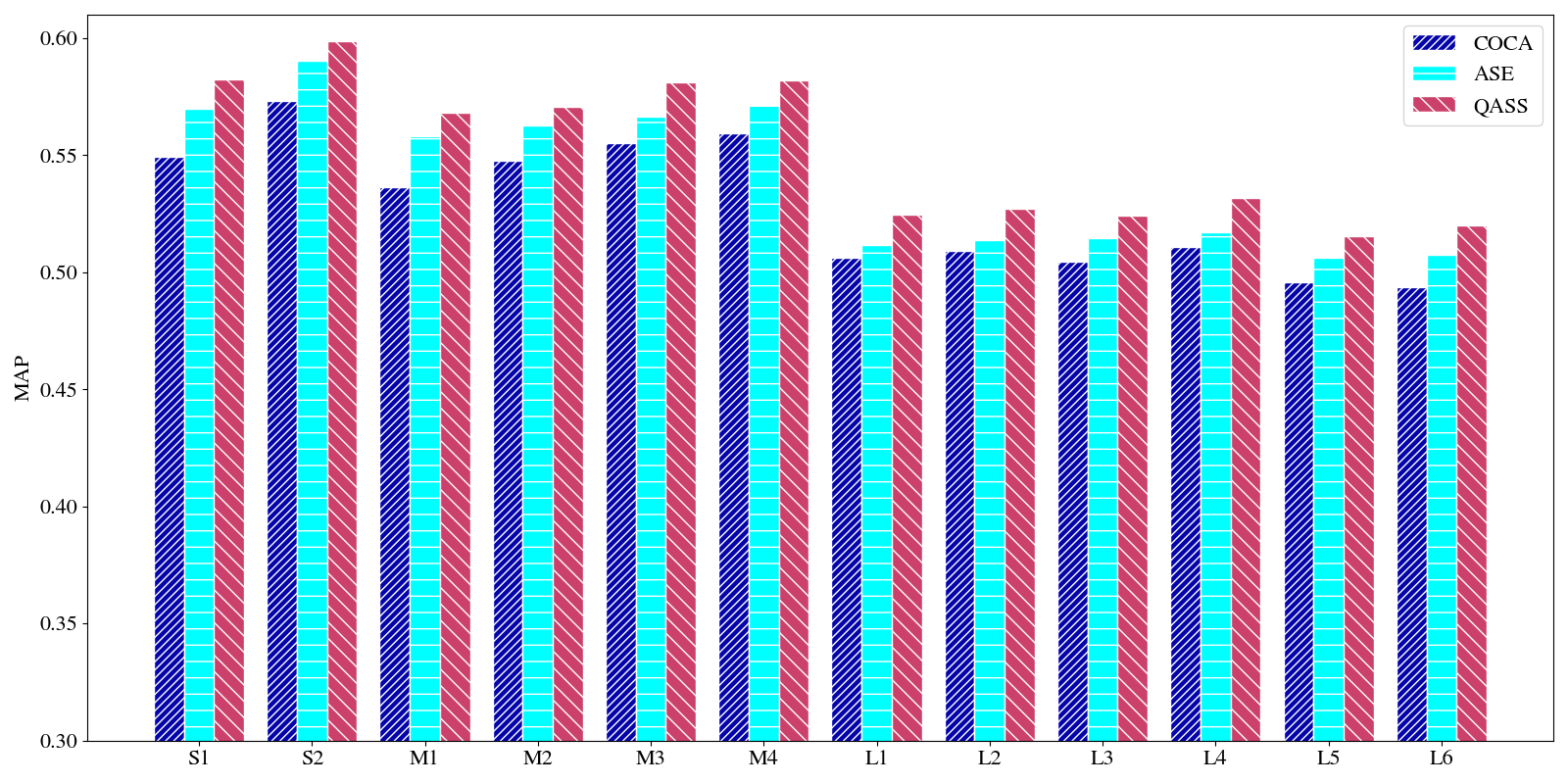}
\caption{We evaluated the performances at different query positions in short (S), medium (M), and long (L) sessions. The numbers appneded to ``S'', ``M'', and ``L'' indicate the query position within the session.}
\label{fig:pos} 
\end{figure}

\subsection{Influence of Hyperparameters}

\subsubsection{The Number of Generated Data Pairs.}
We propose some data augmentation strategies to generate additional data pairs by altering $q_\text{c}$.
We examined how the number of generated data pairs by each strategy (excluding historical queries, which were all used and not treated as a hyperparameter) influenced the results. 
We tuned the numbers within the range of 0 to 10, with increments of 1. 
Due to space constraints, we only present the MAP performance on the AOL dataset and display results for five tuned values. The patterns observed on the other datasets and metrics are similar and thus not included in this discussion.

As illustrated in the left section of Figure~\ref{fig:gen_num}, our data augmentation strategies' performance initially increases to reach optimal values, and then starts to decline.
We believe there is a trade-off: 
If we generate too few data pairs, our model cannot fully model user search patterns. 
However, \ours{} may overfit the generated data if the number is too large.

\subsubsection{Score Margins of Varying Difficulty Training Pairs.}\label{margin_exp}
We propose to coordinate the training of data pairs of varying difficulty by different score margins.
We tune the margins in the range $[0, 1.0]$ with the step of 0.1.
We conduct experiments and present the performance of MRR on the AOL search log to investigate the influence of margins.

As shown in the right part of Figure~\ref{fig:gen_num}, the performances of different margins all increase to optimal values and then drop (except for $m_{\text{rq}}$ which is already set as the maximum value).
There is also a trade-off: 
If the margin is too small, our model will become sensitive and wrongly give high scores for the augmented sequences that are irrelevant to the clicked document.
However, when the margin is too large, our matching model cannot handle strongly relevant distractors.



\subsection{Performances on Different Session Lengths and Query Positions}

Following existing works~\cite{cars, ricr, coca, ase, dcl}, we conduct experiments on different lengths of sessions to analyze our model's performance.
We divide the AOL search log into three kinds of data:

$\bullet$ Short sessions, which consist of 2 queries, account for 66.5\% of the test set.

$\bullet$ Medium sessions, which consist of 3-4 queries, account for 27.24\% of the test set.

$\bullet$ Long sessions, which consist of 5 or more queries, account for 6.26\% of the test set.

We conduct a comparison of the performance of our proposed model, \ours{}, with several baseline models including Duet, COCA, and ASE, using split data. The results presented in Figure~\ref{fig:len} allowed us to make the following observations:
(1)~Context-aware ranking models consistently outperform the ad-hoc ranking model Duet across all lengths of sessions. This finding further supports the notion that incorporating session context is beneficial in understanding user search intents.
(2)~Our model, \ours{}, demonstrates superior performance compared to all other models across all lengths of session data. This result highlights the effectiveness of our proposed query-oriented data augmentation technique.
(3)~All models' performances decline as session lengths become longer, but \ours{} has a smaller drop ratio.
This is because \ours{} constructs additional data pairs by altering queries that have history, which helps our model learn that the same session history can represent different search intents with different current queries.
The results further demonstrate the effectiveness of our model in terms of modeling search patterns of session search.

We also conduct experiments on queries of different positions to study the performance of \ours{} of modeling session progression.
We compared the performance of our model, \ours{}, with COCA and ASE. The results are depicted in Figure~\ref{fig:pos}, and based on these results, we can draw the following observations:
(1)~The performance of all models improved as the session continues. This can be attributed to the availability of more session histories, which further highlights the importance of modeling session context in improving the performance of ranking models.
(2)~\ours{} outperforms baselines at all positions, which demonstrates our query-oriented data augmentation's effectiveness again.
Besides, \ours{} performs especially better at the posterior queries in sessions.
This is because \ours{} emphasizes the modeling of the most important behavior in the session, \ie, the current query, by altering it to construct training pairs, which helps understand search intents when there are lots of behaviors available (long history).
(3)~It is intriguing that all performance drops from L4 to L7.
As stated in~\cite{coca, ase}, these long sessions are believed to be challenging exploratory or extremely complicated search tasks, which are naturally hard to resolve.

\subsection{Cost Analysis}

In this section, we will analyze the cost of \ours{} across three stages: preprocessing, training, and inference.

For the preprocessing stage, most augmentation strategies are rule-based, except for ambiguous query mining. The primary cost here is the time required for mining ambiguous queries, which is approximately 20 minutes for the AOL dataset and 10 minutes for the Tiangong-ST dataset.

For the training stage, the additional time cost in this stage comes from the augmented data pairs. For each original pair, we generate three sequences for term-level modifications, three random queries, four most ambiguous queries, and all historical queries within the session. With an average session length of about three, the average number of sequences generated from historical queries is $(0+1+2)/3=1$ in average. Thus, the total augmented sequences are $3+3+4+1=11$ in average. As a result, QASS requires approximately 11 times the training time compared to the naive BERT model.

For the inference stage, \ours{} incurs the same inference cost as BERT-based models since our augmentation process affects only the training stage. This ensures that \ours{} achieves better performance than BERT-based models while maintaining a similar online cost, which makes it suitable for practical use.

\section{Conclusions and FUTURE WORK}
\label{sec:conclusion}
\IEEEPARstart{I}n this study, we aimed to enhance search logs by incorporating augmented training pairs through query alterations. 
This approach allowed our model to learn that the relevance of a document can vary when the session context changes, thus improving our understanding of users' search patterns. 
The symmetric relevance between the candidate document and the search context was overlooked by the existing training paradigm. 
To generate negative sequences for pair-wise training with the original sequence, we modified the current query at both term and query levels. 
This involved masking, replacing, and adding terms, as well as substituting the query with mined queries from the search log (random queries, historical queries, and ambiguous queries). 
The difficulty of the mined/generated queries varied based on their similarity to the original query. 
Additionally, we employed different score margins to coordinate the data pairs generated from various data augmentation strategies. 
Our approach was evaluated through experiments on two publicly available search logs, and the results demonstrated its effectiveness.

Despite the contributions of our work, there are still some limitations that need to be addressed in future research:
\begin{itemize}
    \item We used term-level modification and query-level replacement to alter the current query.
There are more sophisticated data augmentation strategies to be designed, \eg, masking the word that has the highest attention score.
\item \ours{} was implemented based on BERT. 
However, our approach can be applied to other base models, \eg, the encoder of BART.
In future work, we plan to conduct experiments on different base models to further study our approach's effectiveness and applicability.
\item In this work, we only implemented data augmentation on queries that have session history.
For queries that do not have a history (\ie, ad-hoc queries), we need to design a special augmentation strategy, which may help our model learn more about users' search patterns.
\item For negative training pairs of varying difficulty, we plan to try curriculum learning as a more advanced approach to coordinate them.
\item For representing queries and documents, we will investigate the performance of our augmentation strategies on more advanced general embedding models, such as E5~\cite{wang2022text} or BGE~\cite{bge_embedding}.
\end{itemize}

\ifCLASSOPTIONcompsoc
  \section*{Acknowledgments}
\else
  \section*{Acknowledgment}
\fi

Zhicheng Dou is the corresponding author. This work was supported by the National Natural Science Foundation of China No. 62272467, Beijing Outstanding Young Scientist Program NO. BJJWZYJH012019100020098, Public Computing Cloud, Renmin University of China, and the fund for building world-class universities (disciplines) of Renmin University of China.

\ifCLASSOPTIONcaptionsoff
  \newpage
\fi


\begin{IEEEbiography}[{\includegraphics[width=1in,height=1.25in,clip,keepaspectratio]{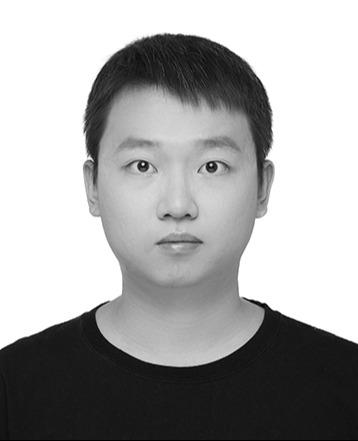}}]{Haonan Chen} received the B.E. degree in computer science and technology from Harbin Institute of Technology, in 2017. And he is studying for Ph.D. at the Gaoling School of Artificial Intelligence, at Renmin University of China. His research interests include information retrieval.
\end{IEEEbiography}

\begin{IEEEbiography}[{\includegraphics[width=1in,height=1.25in,clip,keepaspectratio]{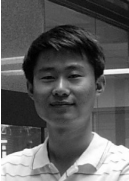}}]{Zhicheng Dou} is currently a professor at Renmin University of China. He received his Ph.D.
and B.S. degrees in computer science and technology from Nankai University in 2008 and 2003, respectively. He worked at Microsoft Research Asia from July 2008 to September 2014. His current research interests are Information Retrieval, Natural Language Processing, and Big Data Analysis. He received the Best Paper Runner-Up Award from SIGIR 2013, and the Best Paper Award from AIRS 2012. He served as the program co-chair of the short paper track for SIGIR 2019. His homepage is http://playbigdata.ruc.edu.cn/dou. He is a member of the IEEE.
\end{IEEEbiography}

\begin{IEEEbiography}[{\includegraphics[width=1in,height=1.25in,clip,keepaspectratio]{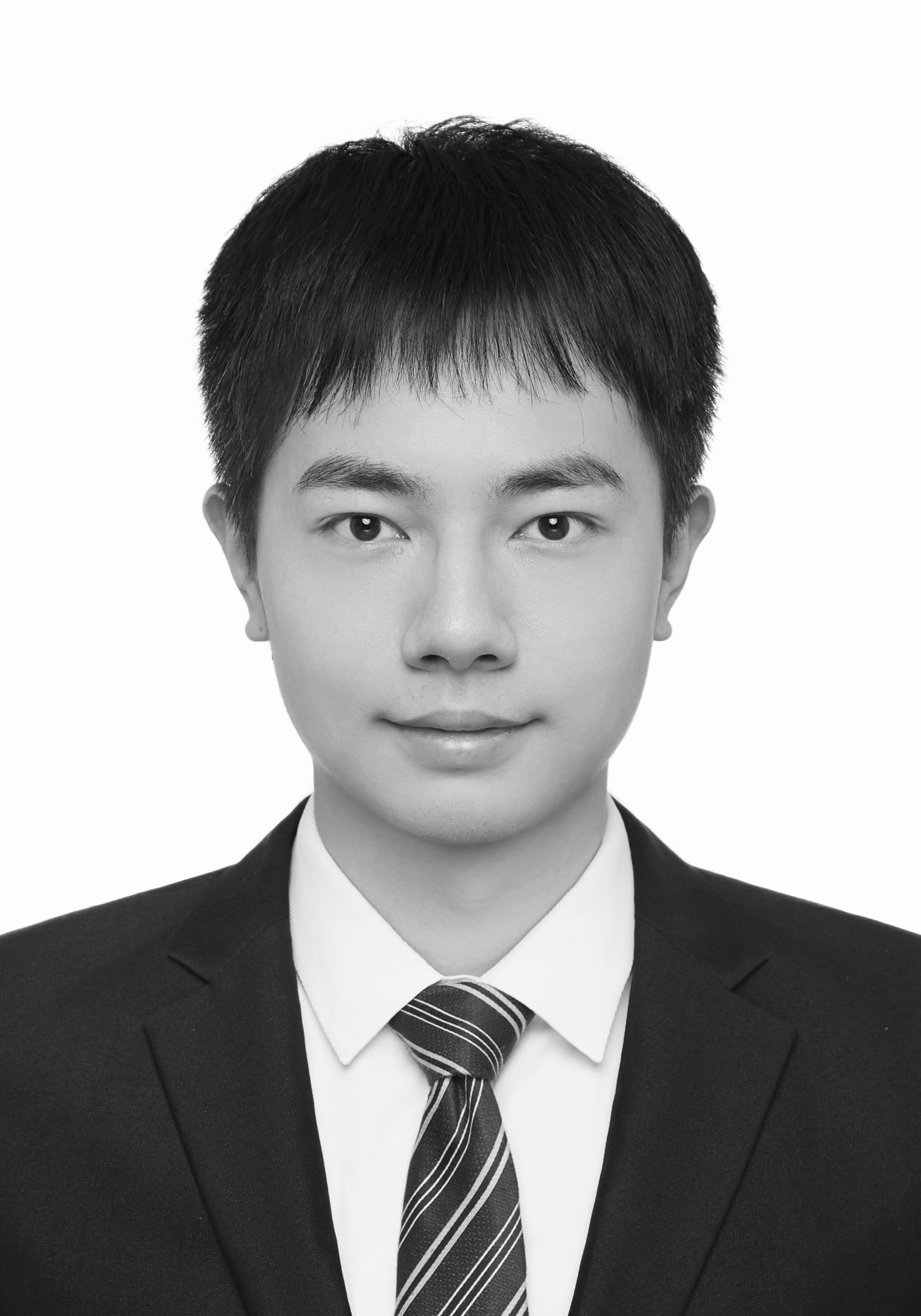}}]{Yutao Zhu} received the B.S. and M.S. degree from Renmin University of China, and the Ph.D. degree from the University of Montreal. He is currently a postdoc at Renmin University of China. His current research interests are Large Language Models and Information Retrieval. He received the Best Paper Award from CCIR 2021 and the Google Scholarship for UdeM in 2019. He served as the PC member of several top-tier conferences, such as ACL, SIGIR, SIGKDD, AAAI, EMNLP, etc. 
\end{IEEEbiography}

\begin{IEEEbiography}[{\includegraphics[width=1in,height=1.25in,clip,keepaspectratio]{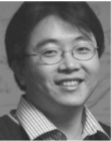}}]{Ji-Rong Wen} received the B,S., and M.S. degrees from Renmin University of China, and the Ph.D. degree from the Chinese Academy of Science, in 1999. He is a professor at the Renmin University of China. He was a senior researcher and research manager with Microsoft Research from 2000 to 2014. His main research interests include web data management, information retrieval (especially web IR), and data mining. He is a senior member of the IEEE.
\end{IEEEbiography}

\end{document}